\documentstyle[seceq,preprint]{ptptex}

 \newcommand{\bref}[1]{(\ref{#1})}

\preprintnumber[3cm]{
NIIG-DP-00-1\\ hep-th/0006180}

\title{
Exact BRS Symmetry realized on the Renormalization Group Flow
}

\author{
Yuji {\sc Igarashi}, Katsumi {\sc Itoh} and Hiroto {\sc So}$^*$
}

\inst{
Faculty of Education, Niigata University, Niigata 950-2181, Japan
\\
$^*$Department of Physics, Niigata University, Niigata 950-2181, Japan
}

\recdate{
}

\abst{Using the average action defined with a continuum analog of the
block spin transformation, we show the presence of gauge symmetry along
the Wilsonian renormalization group flow.  As a reflection of the gauge
symmetry, the average action satisfies the quantum master equation(QME).
We show that the quantum part of the master equation is naturally
understood once the measure contribution under the BRS transformation is
taken into account.  Furthermore an effective BRS transformation
acting on macroscopic fields may be defined from the QME.  The average
action is explicitly evaluated in terms of the saddle point
approximation up to one-loop order.  It is confirmed that the action
satisfies the QME and the flow equation.}

\begin{document}

\maketitle

\section{Introduction}

For the definition of Wilsonian effective
action,\cite{WilsonKogut}\tocite{Polchinski} one
needs to introduce some regularization.  Therefore, it is a nontrivial
problem if symmetries such as chiral or gauge symmetry can survive along
the renormalization group (RG) flow, and if so how they can be realized
in the effective theory.

An important contribution to see a (modified or broken) gauge symmetry
on the RG flow is due to Ellwanger.\cite{Ellwanger} He showed that there
exists the broken Ward-Takahashi (WT) or Slavnov-Taylor identity along
the flow expressed as $\Sigma_k=0$ in his notation,\footnote{We use the
same notation $\Sigma_k$ for the corresponding quantity in our
formulation.} where $k$ denotes an IR cutoff.  Once we find a theory on
the hypersurface defined by $\Sigma_k=0$ in the coupling space, it
remains on the surface along the RG flow and in the limit of $k
\rightarrow 0$ the identity reduces to the Zinn-Justin equation: the
broken WT identity is, in this sense, connected to the usual WT
identity.  This viewpoint suggests that we could modify the gauge
symmetry broken due to the regularization such that it could be
connected smoothly to the usual gauge symmetry.

It had been long believed that the realization of a chiral symmetry
was impossible on the lattice.\cite{NielsenNinomiya}  However L{\"
u}scher\cite{Luescher} took an important step by providing an exact
chiral symmetry on the lattice decade after the Ginsparg-Wilson's
paper.\cite{GW}  His chiral symmetry has a different form compared
with the continuum chiral symmetry.

The above example may suggest the following possibility: a symmetry in a
field theory survives even after a regularization, its form could be
generally different from its familiar form.  In our earlier
publication,\cite{Igarashi} we pursued this possibility in the context
of Wilsonian RG.  We defined a procedure to give an effective field
theory with an IR cutoff.  In this setting it was shown that we may
define a quantity similar to the Ellwanger's $\Sigma_k$: the equation
$\Sigma_k=0$ is found to be the quantum master equation (QME).  We also
constructed explicitly the symmetry transformation on the macroscopic
fields, which was called as the renormalized transformation.  With this
result we claimed that a symmetry survives the regularization and is
kept along the RG flow.  We emphasize that the symmetry on the flow is
exact and it is not ``modified'' or ``broken''.  The Maxwell theory and
the chiral symmetry were the two examples studied in
Ref.~\citen{Igarashi}.  For the latter, we obtained continuum analogs of
the Ginsparg-Wilson relation and the L{\" u}scher's symmetry.

In the present paper we will show that our procedure may be naturally
extended to an interacting gauge theory, typically the non-Abelian gauge
theory coupled to any matter fields.  A major difference from our
earlier examples is the presence of the quantum part in the master
equation.  Although it had been regarded as a ``breaking'' term of the
symmetry, we will see its presence is necessary to keep the symmetry.
The renormalized BRS transformation is given as we did in our previous
paper.  To see more explicitly how our formulation works, we evaluate
the average action with the saddle point approximation up to one-loop
order: it will be shown that the action satisfies both the master
equation and the flow equation.

This paper is organized as follows.  In sect.2, after a brief
explanation of Batalin-Vilkovisky (BV) antifield
formalism,\cite{BV}\footnote{For reviews, see Ref.~\citen{BVreviews}}
the average action is introduced and shown to satisfy the QME and the RG
flow equation.  For the BRS invariance of the average action, the
quantum part of the master equation naturally emerges, which is the
subject of sect.3.  The renormalized BRS transformation is also given.
In sect. 4 we evaluate the average action with the saddle point
approximation.  The last section is devoted to the summary and further
discussions on the average action.  Explanations of our notations will
be found in the Appendix A.  Some relations in sect. 4 are proved in the 
appendices B and C.

Owing to the presence of Grassmann odd fields, we have to keep track of
signs carefully.  In order to make equations correct and, at the same
time, as simple as possible, we will introduce abbreviations whenever
possible.

\section{The average action and its properties}

The average action was introduced by Wetterich\cite{Wetterich} to
realize a continuum analog of the block spin transformation.  Before
presenting it, let us describe the microscopic action and its properties in
the antifield formalism.

\subsection{The antifield formalism}

In the following $\phi_a$ denotes all the fields in the system under
consideration: eg, gauge, ghosts, antighosts, B-fields and matters for
the non-Abelian theory. Further we introduce their antifields
$\phi^*_a$.  For the gauge-fixing, we perform a canonical transformation:
$\phi_{a}\rightarrow \phi_{a},~~\phi_{a}^{*}\rightarrow \phi_{a}^{*}+
\partial \Psi/\partial \phi_{a}$, where $\Psi$ is the gauge fermion,
a function only of the fields. This gauge fixed basis is convenient,
since it retains the antifields. Let $S_{0}[\phi]$ be a BRS invariant
gauge fixed action in the new basis. We consider then an extended
action, linear in the antifields:
\begin{eqnarray} 
S[\phi,\phi^*] \equiv S_{0}[\phi]+\phi^* \delta \phi.
\label{action} 
\end{eqnarray}
Here $\delta \phi_a$ is the BRS transformation of $\phi_a$.  The full
expression of the second term is given in eq.\bref{product}.

Under a set of BRS transformations,
\begin{eqnarray}
\delta \phi_a &=& 
\frac{{\overrightarrow \partial} S}{\partial \phi^*_a}
=(-1)^{\epsilon_a+1}\frac{{\overleftarrow \partial} S}{\partial \phi^*_a},
\nonumber\\
\delta \phi^*_a &=& - \frac{{\overrightarrow \partial} S}{\partial \phi_a}
=(-1)^{\epsilon_a+1}\frac{{\overleftarrow \partial} S}{\partial \phi_a},
\label{BRS tf}
\end{eqnarray}
the extended action  $S[\phi,\phi^*]$ is shown to be invariant:
\begin{eqnarray}
\delta S[\phi,\phi^*] = \delta S_0[\phi] + \phi^* \delta^2 \phi + (-1)^{\epsilon_a+1} \delta \phi_a^* \delta \phi_a=0,
\label{invariance}
\end{eqnarray}
where $\epsilon_a$ is the Grassmann parity of the field $\phi_a$.  The
sign in the third term of eq.\bref{invariance} appears since we have
chosen the BRS transformation to act from the right.  Another important
sign appears in changing a right derivative to a left one and vice
versa, as in \bref{BRS tf}.  See \bref{Sign} for a general formula.

With the antibracket,
\begin{eqnarray}
(F,G)_{\phi} \equiv 
\frac{F {\overleftarrow \partial}}{\partial \phi}
\frac{{\overrightarrow \partial} G}{\partial \phi^*}
-\frac{F {\overleftarrow \partial}}{\partial \phi^*}
\frac{{\overrightarrow \partial} G}{\partial \phi},
\label{antibracket}
\end{eqnarray}
the BRS transformation may be written as $\delta F \equiv (F,S)_{\phi}$.
In terms of the antibracket gauge invariance of the action is nicely
summarized as the master equation: $(S,S)_{\phi}=0$.  In
eq.\bref{antibracket}, the summation over indices and the momentum
integration are implicit.

For the following discussion the action \bref{action} is our starting
point.  So we assume that the action is linear in the antifield
$\phi^*$.  This includes the Yang-Mills fields coupled to matter fields
as a typical and important example.  Actually our consideration may be
extended to an action with nonlinear $\phi^*$ dependence, which will be
discussed in Ref.~\citen{Igarashi2}.

\subsection{The average action}

The average action $\Gamma_k$, with an IR cutoff $k$, is written in
terms of macroscopic fields $\Phi$ after integrating out the high
frequency modes,
\begin{eqnarray}
e^{-\Gamma_{k}[\Phi,\phi^*]/\hbar} &=& 
\int {\cal D} \phi e^{-S_{k}[\phi,\Phi,\phi^*]/\hbar},
\label{averageaction}\\
S_{k}[\phi,\Phi,\phi^*] &=& S_0[\phi] + \phi^* \delta \phi
+ \frac{1}{2}(\Phi - f_{k}\phi)~R_k~(\Phi - f_{k}\phi).
\label{Sk}  
\end{eqnarray}
The third term on the rhs of eq.\bref{Sk} is our abbreviated notation
  for the full expression given in eq.\bref{product2}.  The functions
  $f_k(p)$ and $R_k(p)$ should be chosen appropriately so that the
  macroscopic fields carry momentum less than $k$.  Though
  we do not need their explicit forms in this paper, it would be
  instructive to see how the high frequency modes are integrated out in
  the above path integral.

To realize a continuum analog of the block spin transformation,
Wetterich wrote down some criteria on the functions.  For example,
\begin{eqnarray*}
f_k(p)&=&{\rm exp}\left(-\alpha\left(\frac{p^2}{k^2}\right)^{\beta}\right),\\
\left[R_k(p)\right]_{ab}&=&(1-f_k^2(p))^{-1} \times [{\cal R}_{k}(p)]_{ab} ,
\end{eqnarray*}
with positive $\alpha$ and $\beta$ are the functions which satisfy the
criteria (See Ref.~\citen{Wetterich} for details).  The matrix $[{\cal
R}_{k}(p)]_{ab}$ is at most polynomial in $p$.

Note that: 1) the function $f_k(p)$ is close to one for the momentum
lower than $k$ and decreases rapidly for the higher momentum; 2)
consequently the factor $(1-f_k^2(p))^{-1}$ in $R_k(p)$ is almost
constant for high momentum and getting very large for the momentum lower
than $k$, the $p$ dependence of $[{\cal R}_{k}(p)]_{ab}$ adds only minor
modulation to this behavior.  This implies that $\Phi(p) \sim \phi(p)$
for $p~<~k$, while $\Phi(p)$ with $p~>~k$ does not carry any information
of the microscopic dynamics and appears in a simple quadratic form in
the average action.  In the rest of the paper, we do not need the
functions explicitly and only assume some properties:
$[R_k(p)]_{ab}=(-)^{\epsilon_a \epsilon_b}[R_k(-p)]_{ba}$; the
components of $R_k$ vanish for mixed Grassmann parity indices.

\subsection{The quantum master equation}

An important question is: how the gauge symmetry at the microscopic
level reflects in $\Gamma_k[\Phi,\phi^*]$ ?  The answer was given in our
earlier paper:\cite{Igarashi} the macroscopic action satisfies the QME.

The BRS invariance of the microscopic action may be written as,
\begin{eqnarray}
\int {\cal D} \phi e^{-S_k[\phi+\delta \phi \lambda, \Phi, \phi^*]/\hbar}
- \int {\cal D}\phi e^{-S_k[\phi, \Phi, \phi^*]/\hbar}=0,
\label{BRS inv}
\end{eqnarray}
with the Grassmann odd parameter $\lambda$.  We assumed the BRS
invariance of the measure, ${\cal D}\phi$; thus anomalies are not
considered here.  Rewriting eq.\bref{BRS inv}, we obtain
\begin{eqnarray*}
0 &=& \hbar^2 e^{\Gamma[\Phi,\phi^*]_k/\hbar} \Delta_{\Phi} 
e^{-\Gamma[\Phi,\phi^*]_k/\hbar} 
\equiv \Sigma[\Phi,\phi^*]_k,\\
\Delta_{\Phi} &\equiv& \sum_a (-)^{\epsilon_a+1} \int {\rm d}p f_k(p) \frac{\partial^r}{\partial \Phi_a(-p)}
\frac{\partial^r}{\partial \phi_a^*(p)},
\end{eqnarray*}
or
\begin{eqnarray}
\Sigma_k[\Phi,\phi^*]
=\frac{1}{2}(\Gamma_k[\Phi,\phi^*],\Gamma_k[\Phi,\phi^*])_{\Phi}
- \hbar \Delta_{\Phi} \Gamma_k[\Phi,\phi^*]=0.
\label{Sigma}
\end{eqnarray}
Here the bracket is defined in terms of $\Phi$ and $\phi^*$:
\begin{eqnarray}
(F,G)_{\Phi} \equiv \int {\rm d}^4p f_k(p)[
\frac{F {\overleftarrow \partial}}{\partial \Phi_a(-p)}
\frac{{\overrightarrow \partial} G}{\partial \phi_a^*(p)}
-\frac{F {\overleftarrow \partial}}{\partial \phi_a^*(-p)}
\frac{{\overrightarrow \partial} G}{\partial \Phi_a(p)}].
\label{antibracket2}
\end{eqnarray}
The comparison of eqs.\bref{antibracket} and \bref{antibracket2} suggests 
that $\phi^*/f_k$ may be regarded as the antifield associated with $\Phi$.

\subsection{The flow equation for the average action}

A straightforward calculation leads us to the flow equation:
\begin{eqnarray}
\hspace{-1.2cm}
\hbar \partial_{k}e^{-\Gamma_{k}[\Phi,\phi^*]/\hbar }
&=& - \left[X + \frac{\hbar}{2} {\rm Str}(R_k^{-1}
\partial_{k}R_k) + \hbar~{\rm Str}(\partial_{k}
(\ln f_{k}))\right] e^{-\Gamma_{k}[\Phi,\phi^*]/\hbar}, 
\label{flow eq}\\
X &\equiv&   -\frac{\hbar^2}{2} \frac{\partial^l}{\partial \Phi}
(\partial_{k}R_k^{-1})\frac{\partial^r}{\partial \Phi}+
\partial_{k}(\ln f_{k})\left[\hbar^2 \frac{\partial^l}{\partial \Phi}
R_k^{-1}\frac{\partial^r}{\partial \Phi}+
\hbar \Phi \frac{\partial^l}{\partial \Phi}\right]. 
\label{operator X}
\end{eqnarray}
Here we used the fact, $(R_k)_{\rm even~odd}=(R_k)_{\rm odd~even}=0$,
in our choice for $R_k$.

An interesting property of the quantity $\Sigma_k[\Phi,\phi^*]$ was
found by Ellwanger:\cite{Ellwanger} using the flow equation \bref{flow
eq} we may show the following,
\begin{eqnarray}
\hbar \partial_{k} \Sigma_{k} = (e^{\Gamma_{k}/{\hbar}}X
e^{-\Gamma_{k}/{\hbar}})\Sigma_{k} - e^{\Gamma_{k}/{\hbar}}X~(e^{-\Gamma_{k}/{\hbar}}\Sigma_{k}). 
\label{Ellwanger} 
\end{eqnarray}
Therefore once we are on the hypersurface $\Sigma_k=0$ in the coupling
space, we will keep the same condition even if we change the IR cutoff $k$.

\section{The QME and the renormalized BRS transformation}

In earlier works it had been generally understood that the momentum
cutoff breaks gauge invariance; we only have the condition so that the
gauge invariance recovers when the IR cutoff is removed.  The condition
was beautifully summarized in Ref.~\citen{Ellwanger} and its connection
to the QME was clarified in our earlier paper.\cite{Igarashi} The
commonly shared view is that terms corresponding to $\Delta_{\Phi}
\Gamma_k$ represent the breaking of the gauge invariance.\footnote{If
one uses the average action, the condition is written in a very simple
form as QME.  Of course, in other formalisms it looks completely
different and the ``breaking terms'' look very different in their
appearances.}  Here we show that the BRS invariance will be kept {\it
including} $\Delta_{\Phi} \Gamma_k$ term.

In the following we first explain how a QME is related to the BRS
invariance of a generic gauge invariant system.  One finds the variation
of the path integral measure is exactly the $\Delta_{\Phi} \Gamma_k$
term.  Based on this understanding we may define the renormalized BRS
transformation for the macroscopic fields.

\subsection{A generic gauge system}

Let us consider a generic gauge system with the action ${\cal
A}[\eta,\eta^*]$, where $(\eta,\eta^*)$ could be the microscopic fields
$(\phi,\phi^*)$ or the macroscopic fields $(\Phi,\phi^*)$.
Under the transformation, 
\begin{eqnarray*}  
\eta'&=& \eta + \delta \eta \lambda,\\
\delta \eta &=& (\eta, {\cal A})_{\eta}
= \frac{{\overrightarrow \partial} {\cal A}}{\partial \eta^*},\\
\end{eqnarray*}
we require that the path integral be invariant,
\begin{eqnarray*}
\int {\cal D} \eta e^{-{\cal A}[\eta,\eta^*]/ \hbar}=
\int {\cal D} \eta' e^{-{\cal A}[\eta',\eta^*]/ \hbar},
\end{eqnarray*}
where $\lambda$ is the transformation parameter.  The BRS invariance of
the path integral including the measure may be written as $\delta({\cal
A}[\eta,\eta^*]-\hbar{\rm ln}{\cal D}\eta)=0$.  We will presently see
that this is nothing but a QME and its quantum part is due to the
variation of the measure.  

Let us look at the first term in the above mentioned equation,
\begin{eqnarray*}
\delta {\cal A}[\eta,\eta^*] 
= \frac{{\cal A} {\overleftarrow \partial}}{\partial \eta}\delta \eta
= \frac{{\cal A} {\overleftarrow \partial}}{\partial \eta}
\frac{{\overrightarrow \partial} {\cal A}}{\partial \eta^*}
= \frac{1}{2}({\cal A},{\cal A})_{\eta}.
\end{eqnarray*}
If we assume that the path integral measure is flat, ${\cal D}\eta =
\prod_a {\rm d}\eta_a$, the logarithm of the measure transforms as $\ln
{\cal D}\eta' = \ln {\cal D}\eta + (\delta {\rm ln}{\cal D}{\eta})
\lambda$,\footnote{The argument of eq.\bref{measure} is adapted from
Ref.~\citen{Hata}.}
\begin{eqnarray}
(\delta {\rm ln}{\cal D}{\eta}) \lambda = {\rm ln}~{\rm Sdet}
\frac{\partial^r}{\partial \eta_a}(\eta+\frac{\partial^l {\cal A}}{\partial \eta^*}\lambda)_b \sim \frac{{\overrightarrow \partial}}{\partial \eta^*_a} {\cal A}[\eta,\eta^*]\frac{{\overleftarrow \partial}}{\partial \eta_a} \lambda.
\label{measure}
\end{eqnarray}
Therefore including the contribution from the measure, we obtain the
QME,
\begin{eqnarray}
\frac{1}{2}({\cal A}[\eta,\eta^*],{\cal A}[\eta,\eta^*])_{\eta}
- \hbar \frac{{\overrightarrow \partial}}{\partial \eta^*_a} {\cal A}[\eta,\eta^*]
\frac{{\overleftarrow \partial}}{\partial \eta_a}=0.
\label{mastereq1}
\end{eqnarray}

\subsection{The average action}

Consider the following path integral,
\begin{eqnarray}
&{}& \int {\cal D} \phi e^{-S[\phi,\phi^*]/\hbar} \\
&{}&~~~~~~~~= \int {\cal D} \Phi {\cal D} \phi ~
e^{- {S_{k}[\phi,\Phi,\phi^*]}/\hbar}
= \int {\cal D} \Phi e^{-\Gamma_k[\Phi,\phi^*]/\hbar},
\label{doubleintegral}
\end{eqnarray}
To the original path integral we insert the gaussian integration with
respect to $\Phi$ and reverse the order of the integrations, then we
find the path integral over the average action with the flat measure for
$\Phi$-integration.  The gauge symmetry of the original system is
expressed as the classical master equation.  The path integral of the
average action carries the same information.  As evident from our
general argument, the symmetry is expressed as the QME with its quantum
part $\Delta_{\Phi} \Gamma_k$ coming from the transformation of the path
integral measure.

\subsection{The renormalized BRS transformation}

From the above argument we see that the renormalized BRS transformation
may be read off from the classical part of the QME :\cite{Igarashi}
\begin{eqnarray}
\delta_r \Phi &\equiv& f_k \frac{{\overrightarrow \partial} \Gamma_k}{\partial \phi^*} = f_k \langle \delta \phi \rangle_{\phi},
\label{rBRS}\\
\delta_r \phi^* &\equiv& -f_k \frac{{\overrightarrow \partial} \Gamma_k}{\partial \Phi}
= - \langle f_k R_k(\Phi-f_k \phi) \rangle_{\phi}.
\label{rBRS2}
\end{eqnarray}
Here we used the notation,
\begin{eqnarray}
\left< {\cal O} \right>_{\phi}&=& \int{\cal D}\phi~{\cal O}~e^{-S_k/\hbar} 
\Big/ \int{\cal D}\phi~e^{-S_k/\hbar}.
\end{eqnarray}
In Ref.~\citen{Becchi} the cutoff dependent BRS transformation was
considered in a different approach.

Some comments are in order.  Firstly, let us emphasize that the quantum
part had long been understood to suggest the breaking of the gauge
symmetry, which is not the correct understanding from our viewpoint.
Secondly, as far as we know of, this is the second example where the
quantum part of a QME plays an important role; the first one was the
string field theory(SFT).\cite{Hata}  It is probably very important to
remember the QME is deeply related to the unitarity of the SFT.

\section{The average action in the saddle point approximation}

It would be usually impossible to fully evaluate the path integral
\bref{averageaction} to construct an average action.  In order to
understand the formalism in more concrete terms, a systematic evaluation
of the average action in \bref{averageaction} is definitely instructive.
The loop expansion with the saddle point method suits for our purpose:
it provides a way to integrate out high frequency modes systematically.
In this section we will calculate the average action up to one-loop
order.

The saddle point, $\phi(p)=\phi_0(p)$, is determined by the following
equation,
\begin{eqnarray}
-f_kR_k(\Phi-f_k\phi_0)
+\frac{{\overrightarrow \partial} (\phi^*_{a} P_{a}[\phi_0]+S_{0}[\phi_0])}
{\partial \phi_0}=0,
\label{saddle}
\end{eqnarray}
where $P_{a}[\phi]$ denotes the BRS transformation of
$\phi_{a}$: $P_{a}[\phi] \equiv \delta \phi_{a}$.  The
saddle point equation gives an implicit function,
$\phi_0=\phi_0[\Phi,\phi^*]$.  Note that in eq.\bref{saddle} we have
omitted the indices and the momentum dependence for simplicity.  The left
derivative, ${\overrightarrow \partial} / {\partial \phi_0}$, in the
second term is taken with $\phi^*$ fixed.

Now the average action at the tree level is given as,
\begin{eqnarray}
\Gamma_k^{(0)}[\Phi,\phi^*] \equiv S_k[\phi_0 [\Phi,\phi^*],\Phi,\phi^*],
\label{average0}
\end{eqnarray}
and the one-loop correction is the superdeterminant,
\begin{eqnarray}
{\Gamma_{k}^{(1)}[\Phi,\phi^*]} = 
\frac{\hbar}{2}\ln {\rm Sdet}(A[\phi_0,\phi^*]),
\label{oneloop}
\end{eqnarray}
of the matrix $A$,
\begin{eqnarray}
A_{ab}[\phi_0,\phi^*] = f^2_k [R_k]_{ab} 
+ \frac{\overrightarrow \partial}{\partial \phi_0^a}
(\phi^*_{c}P_{c}[\phi_0] + S_{0}[\phi_0])
\frac{\overleftarrow \partial}{\partial \phi_0^b}.
\label{matrixA}  
\end{eqnarray}

Let us see the one-loop average action, $\Gamma_k^{(0)}+\Gamma_k^{(1)}$,
satisfies both the QME and the flow equation.

\subsection{The one-loop QME}

The QME to be proved may be rewritten as:
\begin{eqnarray}
(\Gamma_k^{(0)},\Gamma_k^{(0)})_{\Phi}&=&0,
\label{master0}\\
(\Gamma_k^{(0)},\Gamma_k^{(1)})_{\Phi}&-&\hbar \Delta_{\Phi} \Gamma_k^{(0)}=0,
\label{master1}
\end{eqnarray}
where we have used the fact, $\frac{1}{2}(\Gamma_k^{(0)},\Gamma_k^{(1)})_{\Phi}
=\frac{1}{2}(\Gamma_k^{(1)},\Gamma_k^{(0)})_{\Phi}$, which is easily seen by using 
${{\overrightarrow \partial} \Gamma_k^{(0)}}/{\partial
\Phi_a}=(-1)^{\epsilon_a}{\Gamma_k^{(0)} {\overleftarrow
\partial}}/{\partial \Phi_a}$ etc.

The tree level master equation \bref{master0} may be confirmed by
using the tree level renormalized BRS transformations for $\Phi$ and $\phi^*$:
\begin{eqnarray}
\delta_r^{(0)} \Phi &=& (\Phi,\Gamma_k^{(0)})_{\Phi}=f_k P[\phi_0]
\label{0rBRS}\\
\delta_r^{(0)} \phi^{*} &=& (\phi^*,\Gamma_k^{(0)})_{\Phi}=-f_k R_k(\Phi-f_k \phi_0)
\nonumber\\
&=& -\frac{{\overrightarrow \partial} 
(\phi^*_{a} P_{a}[\phi_0]+S_{0}[\phi_0])}{\partial \phi_0}.
\label{0rBRS2}
\end{eqnarray}
The final expression in eq.\bref{0rBRS2} follows from the saddle point
equation \bref{saddle}.  Further, using eqs.\bref{0rBRS} and
\bref{0rBRS2}, we may obtain the transformation of the implicit function 
$\phi_0^a[\Phi,\phi^*]$:
\begin{eqnarray}
\delta_r^{(0)}\phi_0^a[\Phi,\phi^*] &=& 
(\phi_0 {\overleftarrow \partial} / \partial \Phi)_{ab} \delta_r^{(0)} \Phi_b
+(\phi_0 {\overleftarrow \partial}  / \partial \phi^*)_{ab} 
\delta_r^{(0)} \phi_b^*\nonumber\\
&=&P_a[\phi_0],
\label{phi0BRS}
\end{eqnarray}
which will be shown in the Appendix B.  From \bref{average0} we see that
$\Gamma_k^{(0)}$ may be written as the rhs of \bref{Sk} with $\phi$
replaced by $\phi_0$.  It is easy to see that the first and second terms
of that expression are invariant under eqs.\bref{0rBRS2} and
\bref{phi0BRS}; the third term of it is also invariant under
eqs.\bref{0rBRS} and \bref{phi0BRS}.  Therefore $\Gamma_k^{(0)}$ is
invariant under \bref{0rBRS}, \bref{0rBRS2} and \bref{phi0BRS}: this
proves the tree level master equation \bref{master0}.

In the Appendix C we will show that the lhs of \bref{master1} reduces to
\begin{eqnarray}
&{ }&(\Gamma_k^{(0)},\Gamma_k^{(1)})_{\Phi}-\hbar \Delta_{\Phi} \Gamma_k^{(0)}
\nonumber\\
&=& \frac{\hbar}{2}(A^{-1})_{ab}
\frac{\overrightarrow \partial}{\partial \phi_0^{b}}
[(\phi^*_{d}P_{d}+S_0)
\frac{\overleftarrow \partial}{\partial \phi_0^{c}}P_c]
\frac{\overleftarrow \partial}{\partial \phi_0^{a}}
-\hbar \frac{\overrightarrow \partial}{\partial \phi^*_a}S[\phi_0,\phi^*]
\frac{\overleftarrow \partial}{\partial \phi_0^a},
\label{oneloopmastereq}
\end{eqnarray}
where $S[\phi_0,\phi^*]$ is the extended action \bref{action} evaluated
at the saddle point.  The first term of eq.\bref{oneloopmastereq}
vanishes owing to the relations,
\begin{eqnarray}
\frac{S_{0}[\phi_0] {\overleftarrow \partial}}
{\partial \phi_0^a} P_a[\phi_0]&=&0,
\label{0BRSS}\\
\frac{P_a[\phi_0] {\overleftarrow \partial}}
{\partial \phi_0^b} P_b[\phi_0]&=&0.
\label{0BRSP}
\end{eqnarray}
These respectively come from the BRS invariance of the action $S_0$
and the nilpotency of the BRS transformation at the microscopic level.
Similarly it is easy to observe that the second term of
eq.\bref{oneloopmastereq} is nothing but the quantum part of the
QME for $S[\phi,\phi^*]$; it vanishes since we assumed that the measure
${\cal D}\phi$ is BRS invariant.

\subsection{The flow equation for the one-loop average action}

Let us see that the one-loop average action satisfies the flow
equation as well.  This is a consistency check of our calculation.
\begin{eqnarray}
-\partial_k \Gamma_k &+& e^{\Gamma_k / \hbar}
[X + \frac{\hbar}{2} {\rm Str}(R_k^{-1}\partial_k R_k)
+\hbar {\rm Str}(\partial_k ({\rm ln}f_k))] e^{- \Gamma_k / \hbar}
\label{oneloopfloweq}\\
&\sim& - \frac{\Gamma_k^{(1)}{\overleftarrow \partial}}
{\partial \phi_0} \partial_k \phi_0
- \frac{\Gamma_k^{(1)}{\overleftarrow \partial}}{\partial \Phi}
[(\partial_k R_k^{-1}) R_k (\Phi-f_k \phi_0)
-\partial_k ({\rm ln}f_k) (\Phi-2f_k \phi_0)].
\nonumber
\end{eqnarray}
The cancellation of O($\hbar^0$) terms follows trivially; thus here on
the rhs we wrote only O($\hbar$) terms.  Remember that
$\Gamma_k^{(1)}$ depends on $\Phi$ only through its $\phi_0$
dependence.  So one may rewrite the $\Phi$ derivative of \bref{oneloopfloweq} into $\phi_0$ derivative; then using \bref{Phiderivative} and
the relation,
\begin{eqnarray}
-\partial_k {f_k} R_k (\Phi-2f_k \phi_0)
-{f_k} \partial_k R_k (\Phi-f_k \phi_0)+ A \partial_k \phi_0=0,
\label{k-derivative of saddle point}
\end{eqnarray}
the vanishing of the rhs of \bref{oneloopfloweq} follows.  The
relation \bref{k-derivative of saddle point} is obtained by
differentiating the saddle point equation.

\section{Summary and Discussions}

By using the average action formalism, we have shown that our claim in
our earlier publication\cite{Igarashi} may be justified even for an
interacting gauge theory: ie, a gauge symmetry survives even with the
presence of a cutoff and the corresponding renormalized BRS
transformation may be constructed from the QME.

The average action satisfies the QME if the original classical action is
gauge invariant.  At this point we have noticed that the antifield
formalism is very convenient to describe the symmetry property of the
average action.  It also follows the flow equation, which also implies
that once the system satisfies the WT identity with some IR cutoff it
will remain so along the RG flow.

The saddle point evaluation is performed for the average action up to
the one-loop order.  The QME and the flow equation are
confirmed explicitly.  As we have seen above, there is no essential
difficulty to extend our analysis to higher orders.  It would be worth
pointing out that the construction of an action satisfying both
equations had not been done earlier.  A related calculation is due to
Ellwanger:\cite{Ellwanger} the gauge mass term was obtained from the
master and flow equations independently and found to be coincide.

The quantum part of a QME had been regarded as an obstacle for the gauge
symmetry.  We have shown that it is necessary for the symmetry since the
measure is not invariant under the renormalized BRS transformation: the
jacobian under the transformation is exactly the quantum part of the
QME.  This argument implies also that we may read off the renormalized
BRS transformation as we did earlier for free field theories.  The
transformation for the averaged field is particularly simple: $\delta_r
\Phi = f_k \langle \delta \phi \rangle_{\phi}$.  Similarly the quantity
$\Sigma_k$ defined referring to the cutoff scale $k$ is also expressed
as a path integral average.  Let us explain briefly how it is so in the
following.

To be observed shortly our argument is applicable even for a microscopic
action with symmetry breaking terms or anomalies.  So let us consider for
the moment the average action $\Gamma_{k}[\Phi,~\Phi^{*}]$ defined with
eq.\bref{averageaction}, but with an action $S[\phi,~\phi^{*}]$ which is
not necessarily BRS invariant. For the microscopic fields, we define the
quantity $\Sigma$ as,
\begin{eqnarray*}
\Sigma[\phi,~\phi^{*}] \equiv \frac{1}{2}\left(S,~S\right)_{\phi} - \hbar \Delta_{\phi} S = {\hbar}^2 
\exp(S/\hbar)\Delta_{\phi}\exp(-S/\hbar).
\label{qmeidentity}
\end{eqnarray*}
The functional average of it may be rewritten as 
\begin{eqnarray}
\left< \Sigma[\phi,~\phi^{*}]\right>_{\phi}&=& \hbar^2 e^{\Gamma_{k}/\hbar} \int{\cal D}\phi~e^{(S-S_k)/\hbar} 
    \left({\Delta_{\phi}}e^{-S/\hbar}\right)\nonumber\\ 
&=& \hbar^2 e^{\Gamma_k/\hbar} \Delta_{\Phi} e^{-\Gamma_k/\hbar} \equiv \Sigma_k[\Phi,\phi^*].
\label{classicalMEtoQME}
\end{eqnarray}
For $S[\phi,~\phi^{*}]$ which does satisfy the (classical) master
equation, eq.\bref{classicalMEtoQME} tells us the average action satisfy
the QME, $\Sigma_k[\Phi,\phi^*]=0$.  This is an important result: the
QME for the average action is obtained from the master equation for the
microscopic action.  Note that the relation \bref{classicalMEtoQME}
holds even for the case that $\Sigma$ does not vanish, which must have
further implications.  For example, it tells us how a symmetry breaking
term changes along the RG flow.

In our formulation, there remain a couple of questions to be clarified.
Among others the following two are particularly important: 1) whether
our QME reduces to the usual Zinn-Justin equation in the limit of $k
\rightarrow 0$; 2) how we prepare the UV theory.  In the forthcoming
paper\cite{Igarashi2} we will show that the approach presented here may
be extended to most general gauge theories.  The relations to other
approaches\cite{Becchi}\tocite{ReuterWetterich} will be given as well;
at the same time it will be explained how the Zinn-Justin equation is
realized in the limit of $k \rightarrow 0$. The second question will be
discussed by introducing an UV cutoff $\Lambda$ and imposing appropriate
boundary conditions on the average action.

\vspace{1cm}
\section*{Acknowledgments}

Discussion with H. Nakano on Ref.~\citen{Hata} is gratefully
acknowledged.

\vspace{1cm}
\appendix
\section{On notations}

The left and right derivatives are written as:
\begin{eqnarray*}
\frac{{\overrightarrow \partial} F}{\partial \phi} 
\equiv \frac{\partial^l F}{\partial \phi},~~~
\frac{F {\overleftarrow \partial}}{\partial \phi}
\equiv \frac{\partial^r F}{\partial \phi}.
\end{eqnarray*}
We find that the notations on the lhs provide us with simpler
expressions for many equations.  However whenever convenient to avoid
possible confusion, we use those on the rhs.

The sign associated with the change from a right derivative to a left
derivative or vice versa is very important,
\begin{eqnarray}
\frac{F {\overleftarrow \partial}}{\partial \chi}
=(-1)^{{\epsilon_{\chi}}(\epsilon_F+1)}
\frac{{\overrightarrow \partial} F}{\partial \chi}.
\label{Sign}
\end{eqnarray}

Here we explain our abbreviated notations for some examples.  The second
term of $S[\phi,\phi^*] \equiv S_{0}[\phi]+\phi^* \delta \phi$ is the
shorthand notation for
\begin{eqnarray}
\phi^* \delta \phi \equiv \sum_a \int {\rm d}^4p \phi^*_a(-p) \delta \phi_a(p).
\label{product}
\end{eqnarray}
In the multiplication on the lhs the summation over the index $a$ and
the momentum integration are implicit.  Similarly in the block spin
transformation we use the following,
\begin{eqnarray}
(\Phi-f_k \phi) R_k (\Phi-f_k \phi) \equiv 
\int {\rm d}^4p(\Phi-f_k \phi)^a(-p) [R_k(p)]_{ab} (\Phi-f_k \phi)^b(p).
\label{product2}
\end{eqnarray}

\section{A proof of eq.\bref{phi0BRS}: $\delta_r^{(0)}\phi_0=P[\phi_0]$}

Here we show the following equation:
\begin{eqnarray*}
P[\phi_0]=(\partial^r \phi_0 / \partial \Phi) \delta_r^{(0)} \Phi 
+ (\partial^r \phi_0 / \partial \phi^*) \delta_r^{(0)} \phi^*.
\end{eqnarray*}

By differentiating the saddle point equation \bref{saddle} with respect to
$\Phi$ and $\phi^*$, we obtain the relations,
\begin{eqnarray}
f_kR_k &=& A (\partial^r \phi_0 / \partial \Phi),
\label{Phiderivative}\\
\frac{\overrightarrow \partial}{\partial \phi_0} (\phi^*_{a} P_{a})
\frac{\overleftarrow \partial}{\partial \phi^*}\Big|_{\phi_0}
  &+& A (\partial^r \phi_0 / \partial \phi^*)=0,
\label{phistarderivative}
\end{eqnarray}
where $A_{ab}[\phi_0,\phi^*]$ is defined in eq.\bref{matrixA}.  
In eq.\bref{phistarderivative}, the $\phi^*$ derivative in the first
term is taken with $\phi_0$ fixed, which is denoted by the subscript
$\phi_0$.

Using them and the tree level renormalized BRS transformation, the
equation to be proved may be rewritten as,
\begin{eqnarray*}
P=A^{-1} f_k^2 R_k P + A^{-1} 
\left[\frac{\overrightarrow \partial}{\partial \phi_0}(\phi^*_{a} P_{a})
\frac{\overleftarrow \partial}{\partial \phi^*}\Big|_{\phi_0}\right]
\frac{\overrightarrow \partial}{\partial \phi_0}(\phi^*_{b} P_{b}+S_0).
\end{eqnarray*}
Let us see the vanishing of the difference of lhs and rhs multiplied by $A$,
\begin{eqnarray}
\hspace{-1cm}
&{ }&(A-f_k^2R_k)_{ab}P_b - 
\left[\frac{\overrightarrow \partial}{\partial \phi_0}(\phi^*_{c} P_{c})
\frac{\overleftarrow \partial}{\partial \phi^*}\Big|_{\phi_0}\right]_{ab}
\frac{\overrightarrow \partial}{\partial \phi_0^b}(\phi^*_{d} P_{d}+S_0)
\nonumber\\
\hspace{-1cm}
&=& \left[\frac{\overrightarrow \partial}{\partial \phi_0}
(\phi^*_{c} P_{c}+S_0)
\frac{\overleftarrow \partial}{\partial \phi_0}\right]_{ab}P_b
-
\left[\frac{\overrightarrow \partial}{\partial \phi_0}(\phi^*_{c} P_{c})
\frac{\overleftarrow \partial}{\partial \phi^*}\Big|_{\phi_0}\right]_{ab}
\frac{\overrightarrow \partial}{\partial \phi_0^b}(\phi^*_{d} P_{d}+S_0),
\label{equation}
\end{eqnarray}
where we substituted eq.\bref{matrixA} on the rhs.  Taking the
$\phi^*$-differentiation, we rewrite the second term;
\begin{eqnarray*}
&-&
\left[\frac{\overrightarrow \partial}{\partial \phi_0}(\phi^*_{c} P_{c})
\frac{\overleftarrow \partial}{\partial \phi^*}\Big|_{\phi_0}\right]_{ab}
\frac{\overrightarrow \partial}{\partial \phi_0^b}(\phi^*_{d} P_{d}+S_0)\\
=
&-&
\left( \frac{\overrightarrow \partial}{\partial \phi_0^a}
P_b (-)^{\epsilon_b+1} \right)
\frac{\overrightarrow \partial}{\partial \phi_0^b}(\phi^*_{d} P_{d}+S_0)\\
=
&{ }&
\left( \frac{{\overrightarrow \partial}P_b}{\partial \phi_0^a} \right)
\left[ (\phi^*_{d} P_{d}+S_0)
\frac{\overleftarrow \partial}{\partial \phi_0^b} \right]
= (-)^{\epsilon_a \epsilon_b}
\left[ (\phi^*_{d} P_{d}+S_0)
\frac{\overleftarrow \partial}{\partial \phi_0^b} \right]
\left( \frac{{\overrightarrow \partial}P_b}{\partial \phi_0^a} \right).
\end{eqnarray*}
Thus the rhs of \bref{equation} may be rewritten as, 
\begin{eqnarray*}
\frac{\overrightarrow \partial}{\partial \phi_0^a}
\left( [\phi^*_{c} P_{c}+S_0] 
\frac{ \overleftarrow \partial}{\partial \phi_0^{b}} P_{b}\right),
\end{eqnarray*}
which vanishes owing to eqs.\bref{0BRSS} and \bref{0BRSP}.

\section{A proof of eq.\bref{oneloopmastereq}: the QME to one-loop order}

In \bref{oneloopmastereq} the first term is the variation of
$\Gamma_k^{(1)}$ by the tree level BRS transformation given in
\bref{0rBRS} and \bref{0rBRS2}:
\begin{eqnarray}
&(&\Gamma_k^{(0)},\Gamma_k^{(1)})_{\Phi}- \hbar \Delta_{\Phi} \Gamma_k^{(0)}
\nonumber\\
&=& \frac{\hbar}{2}{\rm Str} A^{-1} \left(          
\left( \frac{A {\overleftarrow \partial}}{\partial \phi^*_{a}}\Big|_{\phi_0} \right)
\delta_r^{(0)}\phi^*_{a}+
\left( \frac{A {\overleftarrow \partial}}{\partial \phi^{a}_0} \right)
\delta_r^{(0)}\phi^{a}_0
\right)
- \hbar {\rm tr} (f_k \partial^r P / \partial \Phi).
\label{master equation}
\end{eqnarray}
Since the matrix $A$ is a function of $\phi_0$ and $\phi^*$, the
variation under the tree level BRS transformation is taken with
respect to those variables.  The derivatives in the first term of
\bref{master equation} should be understood accordingly.  The second
term is the trace (not the supertrace) of the matrix $\partial^r P_a /
\partial \Phi_b$, which may be rewritten by using eqs. \bref{matrixA}
and \bref{Phiderivative} as
\begin{eqnarray*}
\Delta_{\Phi}\Gamma_k^{(0)} &=& 
{\rm tr} (f_k \partial^r P / \partial \phi_0 \cdot \partial^r \phi_0 / \partial \Phi))\\
&=& {\rm tr} (\partial^r P / \partial \phi_0 A^{-1} f_k^2 R_k)\\
&=& {\rm tr} \left(\partial^r P / \partial \phi_0 A^{-1} [A- \frac{\overrightarrow \partial}{\partial \phi_0}(\phi^*_c P_c+S_0)\frac{\overleftarrow \partial}{\partial \phi_0}]\right).
\end{eqnarray*}
Therefore eq.\bref{master equation} becomes,
\begin{eqnarray*}
(\Gamma_k^{(0)},\Gamma_k^{(1)})_{\Phi}&-& \hbar \Delta_{\Phi} \Gamma_k^{(0)}\\
&=& - \hbar {\rm tr}(\partial^r P / \partial \phi_0)+
\frac{\hbar}{2}{\rm Str} A^{-1} \left(          
-\left( \frac{A {\overleftarrow \partial}}{\partial \phi^*_a}\Big|_{\phi_0} \right)
\frac{{\overrightarrow \partial}}{\partial \phi^a_0}
(\phi^*_{c} P_{c} + S_0)+
\left( \frac{A {\overleftarrow \partial}}{\partial \phi^a_0} \right)
P_{a}
\right)\\
&+& \hbar~{\rm tr}\left[A^{-1} 
\left( \frac{{\overrightarrow \partial}}{\partial \phi_0}
(\phi^*_{c} P_{c} + S_0)
\frac{{\overleftarrow \partial}}{\partial \phi_0}\right)
\partial^r P / \partial \phi_0\right].
\end{eqnarray*}
We may write the rhs more explicitly.  After the
$\phi^*$-differentiation, the second and third terms are written as
\begin{eqnarray*}
&{ }&\frac{\hbar}{2} (-)^{\epsilon_a}A^{-1}_{ab} \left(          
- (-)^{(\epsilon_{c}+1)(\epsilon_a+1)}
\left( \frac{{\overrightarrow \partial}}{\partial \phi_0^b} P_{c}  
\frac{{\overleftarrow \partial}}{\partial \phi_0^a} \right)
\frac{{\overrightarrow \partial}}{\partial \phi^{c}_0}
(\phi^*_{d} P_{d} + S_0)+
\left( \frac{A_{ba} {\overleftarrow \partial}}{\partial \phi^{c}_0} \right)
P_{c} \right)\\
&+& \hbar A^{-1}_{ab} 
\left( \frac{{\overrightarrow \partial}}{\partial \phi_0^b}
(\phi^*_{d} P_{d} + S_0)
\frac{{\overleftarrow \partial}}{\partial \phi_0^c} \right)
\left( \frac{P_c {\overleftarrow \partial}}{\partial \phi_0^a} \right).
\end{eqnarray*}
An easy calculation leads us to eq.\bref{oneloopmastereq}: one must
take care of signs carefully, in particular, those coming from
eq.\bref{Sign}.


\vspace{0.5cm}
\begin{thebibliography}{99}
\bibliographystyle{unsrt} 

%
\setlength{\itemsep}{0.0in}

\bibitem{WilsonKogut} K. G. Wilson and J. Kogut,
Phys. Rep. {\bf C12} (1974) 75.

\bibitem{WegnerHoughton} F. J. Wegner and A. Houghton,
Phys. Rev. {\bf A8} (1973) 401.

\bibitem{Polchinski} J. Polchinski, 
Nucl. Phys. {\bf B231} (1984) 269.

\bibitem{Ellwanger} U. Ellwanger,
Phys. Lett. {\bf B335} (1994) 364.

\bibitem{NielsenNinomiya} H. Nielsen and M. Ninomiya, 
Nucl. Phys. {\bf B185} (1981) 20, ERRATUM-ibid.{\bf B195} (1982) 541;
ibid {\bf B193} (1981) 173;  Phys. Lett. {\bf 105B} (1981) 219.

\bibitem{Luescher} M. L{\" u}scher,
Phys. Lett. {\bf B428} (1998) 342; Nucl. Phys. {\bf B549} (1999) 295.

\bibitem{GW} P. Ginsparg and K. Wilson, 
Phys. Rev. {\bf D25} (1982) 2649.

\bibitem{Igarashi}Y. Igarashi, K. Itoh and H. So,
Phys. Lett. {\bf B479} (2000) 336, hep-th/9912262.

\bibitem{BV} I. A. Batalin and G. A. Vilkovisky, Phys. Lett. {\bf 102B} (1981) 27.

\bibitem{BVreviews} M. Henneaux and C. Teitelboim, {\it Quantization of
		Gauge Systems}, (1992) Princeton University Press;
J. Gomis, J. Paris and S. Samuel, Phys. Rept. {\bf 259} (1995) 1-145;
W. Troost and A. Van Proeyen, {\it An introduction to Batalin-Vilkovisky
		Lagrangian Quantisation}, unpublished notes.

\bibitem{Wetterich} C. Wetterich,
Nucl. Phys.{\bf B352} (1991) 529; Z. Phys. {\bf C60} (1993) 461.

\bibitem{Hata} H. Hata, Nucl. Phys. {\bf B329} (1990) 698.

\bibitem{Igarashi2}Y. Igarashi, K. Itoh and H. So, in preparation.

\bibitem{Becchi} C. Becchi,
{On the construction of renormalized quantum field theory using
		renormalization group techiniques}, in {\it Elementary
		particles, Field theory and Statistical mechanics}, eds. M. Bonini, G. Marchesini and E. Onofri, Parma University 1993.

\bibitem{Bonini} M. Bonini, M. D'Attanasio and G. Marchesini,
		Nucl. Phys. {\bf B418} (1994) 81; {\it ibid} {\bf B421}
		(1994) 429; {\it ibid} {\bf B437} (1995) 163;
Phys. Lett. {\bf B346} (1995) 87; M. Bonini and G. Marchesini,
		Phys. Lett. {\bf B389} (1996) 566.

\bibitem{DAttanasio} M. D'Attanasio and T. R. Morris, 
Phys. Lett.  {\bf B378} (1996) 213.

\bibitem{ReuterWetterich} M. Reuter and C. Wetterich,
Nucl. Phys. {\bf B 417} (1994) 181; {\it ibid} {\bf B 427} (1994) 291;
F. Freire and C. Wetterich, Phys. Lett. {\bf B380} (1996) 337.

\end{thebibliography}
\end{document}